\documentstyle[amssymb]{article}
\newtheorem{defn}{Definition}

\newtheorem{rem}{Remark}

\newtheorem{cor}{Corollary}
\newtheorem{prop}{Proposition}
\newtheorem{theo}{Theorem}

\newcommand{\Rn}{{\rm I\!R}} 
\newcommand{\Cn}{{\setbox0=\hbox{
$\displaystyle\rm C$}\hbox{\hbox
to0pt{\kern0.6\wd0\vrule height0.9\ht0\hss}\box0}}} 
\newcommand{\idty}{\hat{\rm 1\mskip-4mu l}} 
\newcommand{\Zn}{{\hbox{$\sf\textstyle Z\kern-0.3em Z$}}} 

\newcommand{\pr}{{\bf Proof}:\quad}
\newcommand{\QED}{\\ \null \hfill $\Box$}

\newcommand{\de}{\delta}

\newcommand{\la}{\lambda}
\newcommand{\om}{\omega}

\newcommand{\cA}{{\cal A}}
\newcommand{\cB}{{\cal B}}

\newcommand{\cE}{{\cal E}}
\newcommand{\cF}{{\cal F}}
\newcommand{\cH}{{\cal H}}
\newcommand{\cI}{{\cal I}}

\newcommand{\cM}{{\cal M}}
\newcommand{\cN}{{\cal N}}

\newcommand{\cS}{{\cal S}}

\begin{document}
\title{On entanglement of states and quantum correlations
    \thanks{ 
\ \ Work supported by KBN grant PB/0273/PO3/99/16 
}}
\author{
W\ A\ Majewski 
    \thanks{\ \ Institute of Theoretical Physics and Astrophysics, University
of Gda\'nsk, Wita Stwosza 57, PL 80-952 Gda\'nsk, Poland.  E-mail: fizwam@univ.gda.pl}
}
\date{}
\maketitle{}


\begin{abstract}
In this paper we present the novel qualities of entanglement
of formation for general (so also infinite dimensional)
quantum systems and we introduce the notion of coefficient of quantum correlations.
Our
presentation stems from rigorous description of entanglement of formation. 
\end{abstract}

\section{Introduction}
The problem of quantum entanglement of mixed states has attracted
much attention recently and that concept has been widely considered in different 
physical contexts (cf. \cite{per}, \cite{Hor1} and references therein, 
see also \cite{VW}, \cite{Woo}, \cite{VP}, \cite{KM}). 
Moreover, it is frequently argued that the nature of entangled states is strongly 
related to quantum correlations.

In this paper we are concerned with the generalization of the entanglement of formation,
introduced in \cite{Ben} as well as with the rigorous definition of a measure of 
quantum correlations. To this end, firstly  
 we look more closely at the original definition of EoF. Namely,
there is a difficulty in implementing the definition given by 
Bennett {\it et al}
in the sense that it is not clear why
the operation of taking $min$ over the set of
all decomposition of the given state into 
finite convex combination of pure states
is well defined (for details see \cite{M}). To overcome this problem
and to get a measure with nice properties we shall
use the theory of decomposition which is based on the theory
of compact convex sets and boundary integrals. Then, having rigorously described 
measure of entanglement we will discuss the concept of coefficient of quantum correlations.
The paper is organized as follows. In Section II we set up notation
and terminology, and we review some of the standard facts on the theory
of decomposition. Section III contains our definition of entanglement of 
formation, EoF, with theorem 1 saying that
 EoF is equal to zero if and
only if the state is a separable one. 
In section IV we review properties of EoF.
In the final section V, we present the concept of coefficient of quantum correlations
with a discussion of its relations to entanglement.

\section{Preliminaries}

Let us consider a composite system $"1+2"$ and its Hilbert space
of pure states ${\cal H}_1 \otimes {\cal H}_2$ 
where ${\cal H}_i$ is the Hilbert space associated to subsystem
$i$ ($i = 1,2$). Let ${\cal B}({\cal H}) $
denote the set of all bounded
linear operators on ${\cal H}$. Unless otherwise stated, $\cM$ stands for a
(unital) $C^*$-subalgebra of 
$\cB(\cH_2)$.
We will assume that $\cH_1$ is a finite dimensional
space (for a discussion how to dispense
with that assumption see \cite{M}).
$\cH_2$ will be an arbitrary (infinite dimensional,
separable) Hilbert space.
In other words, the composite system consists of small subsystem and 
a big heat-bath, rather a typical situation for concrete physical
problems. 

Turning to states we recall that
any density matrix (positive operator of trace equal to $1$) on $\cal H$ determines 
uniquely a linear
positive, normalized, functional $\omega_{\varrho}(\cdot) \equiv \omega (\cdot)
\equiv Tr\{ \varrho \cdot \}$ on ${\cal B}({\cal H})$ which is also called 
a normal state.
We will assume Ruelle's separability condition for $\cM$ (cf. \cite{Ru1},
\cite{Ru2}, \cite{BR}): a subset $\cF$ of the set of all states $\cS$ 
of $\cM$ satisfies Ruelle's separability [Note: this refers to topological
properties, and is not related to the algebraic notion, which is the subject of this paper]
condition if there exists a sequence
$\{ \cM_n \}$ of sub-$C^*$-algebras of $\cM$ such that 
$\cup_{n \ge 1} \cM_n$ is dense in $\cM$, and each $\cM_n$ contains
a closed, two-sided, separable ideal $\cI_n$ such that
\begin{equation}
\cF = \{ \omega; \omega \in \cS, ||\omega|_{\cI_n} || = 1, n\ge 1 \}
\end{equation}

We recall that this condition leads to a situation 
in which the subsets of states have good measurability properties (cf \cite{BR}). 
Furthermore, one can easily verify that this separability
condition is satisfied in our case provided that we restrict to the set of 
normal states on $\cM$ or $\cM$ is a separable $C^*$-algebra.

The density matrix $\varrho$ (state) on the Hilbert space 
${\cal H}_1 \otimes {\cal H}_2$ is called separable if it can be written or 
approximated (in the norm) by the density matrices (states)
of the form: 
$$\varrho = \sum p_i \varrho_i^1 \otimes \varrho_i^2 \qquad 
\Biggl( \omega(\cdot)
= \sum p_i (\omega^1_i \otimes \omega^2_i) ( \cdot) \Biggr)$$
where $p_i \ge 0$, $\sum_i p_i = 1$, $\varrho^{\alpha}_i$ are density
matrices on ${\cal H}_{\alpha}$, $\alpha = 1,2$, and
$(\omega_i^1 \otimes \omega_i^2)(A \otimes B)
\equiv \omega^1_i(A) \cdot \omega^2_i(B) \equiv (Tr \varrho_i^1 A) 
\cdot (Tr \varrho_i^2 B) \equiv Tr\{ \varrho_i^1 \otimes \varrho_i^2 \cdot
A \otimes B\}$.
In other words, separable states are the norm-closed convex hull of all 
product states on $\cB(\cH_1) \otimes \cM$ (or more generally, on the 
tensor product of two $C^*$-algebras). It is well known ( see e.g. \cite{Kad}) the 
state space of the tensor product $\cA_1 \otimes \cA_2$ ($\cN_1 \overline{\otimes}
\cN_2$) of two $C^*$-algebras $\cA_1$ and $\cA_2$ (two $W^*$-algebras $\cN_1$ and $\cN_2$
respectively) is not the norm-closed (weak$^*$-closed) convex hull of all
product states on $\cA_1 \otimes \cA_2$ ($\cN_1 \overline{\otimes} \cN_2$).
Thus, one can define

\begin{defn}
Non- separable states are called entangled states. The set of entangled states
is defined by
\begin{equation}
\cS_{entangled} \equiv \cS_{en} = \cS \setminus \{separable \quad states \}
\end{equation}
where $\cS$ stands for the state space.
\end{defn}
\smallskip

Now, for the convenience of the reader, we introduce some terminology
and give a short resum\'e of results from convexity and Choquet
theory that we shall need in the sequel
(for details see \cite{phel}, \cite{skau}, \cite{Mey}, and \cite{BR}). Let 
$\cA$ stand for a $C^*$-algebra. From now on we make the same assumption
of Ruelle separability for $\cA$ which was posed for $\cM$. In next sections, by a 
slight abuse of notation we will write $\cA$ for $\cB(\cH_1) \otimes \cM$.
By $\cS$ we will denote the state space of 
$\cA$, i.e. the set of linear, positive, normalized,
 linear functionals on $\cA$.
We recall that $\cS$ is a compact convex set in the $^*$-weak topology.
Further, we denote by $M_1(\cS)$ the set of all probability
Radon measures on $\cS$. It is well known that
$M_1(\cS)$ is a compact subset of the vector space of real, regular
Borel measures on $\cS$. Further, let us recall
the concept of barycenter $b(\mu)$ of a measure $\mu \in M_1(\cS)$:
\begin{equation}
b(\mu) = \int d\mu (\varphi) \varphi
\end{equation}
where the integral is understood in the weak sense. The set
$M_{\omega}(\cS)$ is defined as a subset of $M_1(\cS)$
with barycenter $\omega$, i.e.
\begin{equation}
M_{\omega}(\cS) = \{ \mu \in M_1(\cS), b(\mu) = \omega \}
\end{equation}
$M_{\omega}(\cS)$ is a convex closed subset of $M_1(\cS)$, hence
compact in the weak $^*$-topology.
Thus, it follows by the Krein-Milman theorem that there are "many"
 extreme points in $M_{\omega}(\cS)$. We say 
the measure $\mu$ is simplicial
if $\mu$ is an extreme point in $M_{\omega}(\cS)$. The set of all simplicial 
measures in $M_{\omega}(\cS)$ will be denoted by $\cE(\cS)$.

\section{Entanglement of Formation}

Let us define, for a state $\omega$ on ${\cal B}({\cal H}_1) \otimes
\cM$ the following map:
\begin{equation}
\label{r}
(r \omega)(A) \equiv \omega(A \otimes {\bf 1})
\end{equation}
where $A \in {\cal B}({\cal H}_1)$.

Clearly, $r \omega$ is a state on ${\cal B}({\cal H}_1)$. One has

\smallskip

{\it Let $(r \omega)$ be a pure state on ${\cal B}({\cal H}_1)$
(so a state determined by a vector
from ${\cal H}_1$). Then $\omega$ can be written as a product state on
${\cal B}({\cal H}_1) \otimes \cM$.}

The proof of that statement can be extracted from \cite{tak}. 
(For more details we refer the reader to \cite{tak}, \cite{MM}, \cite{M}).

Conversely, there is another result in operator 
algebras saying that if $\omega$ 
is a state on ${\cal B}({\cal H}_1)$ then there exists a state
$\omega^{\prime}$ over ${\cal B}({\cal H}_1) \otimes \cM$
which extends $\omega$. If $\omega$ is a pure state of
${\cal B}({\cal H}_1)$ then $\omega^{\prime}$ may be chosen
to be a pure state of ${\cal B}({\cal H}_1) \otimes \cM$
(cf. \cite{BR}). This observation is the crucial one for
our definition of entanglement of formation which is phrased in terms of decomposition
theory.

\begin{defn}

Let $\omega$ be a state on ${\cal B}({\cal H}_1) \otimes \cM$. 
{\it The entanglement of formation, EoF, is
 defined
as
\begin{equation}
{E}(\omega) = inf_{\mu \in M_{\omega}(\cS)}
\int_{\cS} d\mu(\varphi) S(r\varphi)
\end{equation}
where $S(\cdot)$ stands for the von Neumann entropy, i.e. $S(\varphi)
= - Tr \varrho_{\varphi} log \varrho_{\varphi}$
 where $\varrho_{\varphi}$
is the density matrix determining the state $\varphi$.}
\end{defn}
\vskip 1cm

To comment the above definition we recall that the map $r$ 
and the function $S$ are ($^*$-weakly ) continuous.
At this point we want to strongly emphasize that we use 
the entropy function $S$ only to respect tradition. Namely, to have
a well defined concept of EoF we need a concave 
non-negative continuous function 
which vanishes on pure states (and only on pure states).
In our case, with the first subsysten being finite, the von Neumann entropy
meets these conditions.
Clearly, there are others functions satisfying these conditions.
Our next remark is that 
we define EoF as infimum of integrals evaluated on continuous
function and the infimum is taken over a compact set. Therefore,
the infimum is attainable, i.e. there exists a measure
$\mu_0 \in M_{\omega}(\cS)$ such that
\begin{equation}
E(\omega) = \int_{\cS} d\mu_0(\varphi)S(r \varphi)
\end{equation}
and
\begin{equation}
\omega = \int_{\cS} d\mu_0(\varphi)\varphi
\end{equation}

\vskip 1cm 
To argue that $E(\omega)$ is a well defined measure of entanglement one should show
that $\cF \ni \omega \mapsto E(\omega) $ 
is equal to $0$ 
only for separable states (we recall that $\cF$ stands for the subset 
of states satisfying Ruelle's condition, cf. Section II). This is the case. Namely, 
one can prove (see \cite{M})
\begin{theo}
A state $\omega \in \cF$ is separable if and only if EoF $E(\omega)$
is equal to 0.
\end{theo}

\section{Properties of EoF}
In this section we list briefly properties of EoF. We start with 
\subsection{Convexity of EoF}

Firstly, let us observe that the set
$M_{\lambda_1 \omega_1 + \lambda_2 \omega_2}(\cS)$
contains the sum of the sets $\lambda_1 M_{\omega_1}(\cS)$
and $\lambda_2 M_{\omega_2}(\cS)$ where $\lambda_1$ and $\lambda_2$
are non-negative numbers such that $\lambda_1 + \lambda_2 =1$.
To see this we recall (see e.g. \cite{BR} or \cite{Alf})
that $\mu \in M_{\omega}(\cS)$ if and only if $\mu(f) \ge f(\omega)$
for any continuous, real-valued, convex function $f$.
Thus
\begin{equation}
(\lambda_1 \mu_1 +\lambda_2 \mu_2)(f) \ge \lambda_1 f(\omega_1)
+\lambda_2 f(\omega_2) \ge f(\lambda_1 \omega_1 + \lambda_2 \omega_2)
\end{equation}
implies the above stated relation between sets. 
Hence
\begin{eqnarray}
E(\lambda_1 \omega_1 + \lambda_2 \omega_2)
= \inf_{\mu \in M_{\lambda_1 \omega_1 + \lambda_2 \omega_2}(\cS)}
\int d\mu(\varphi) S(r\varphi) \nonumber \\ 
\le \lambda_1
\inf_{\mu \in M_{\omega_1}(\cS)} \int d\mu(\varphi) S(r \varphi) \nonumber \\
+ \lambda_2 \inf_{\mu \in M_{\omega_2}(\cS)} \int d\mu(\varphi)
S(r \varphi) 
 = \lambda_1 E(\omega_1) + \lambda_2 E(\omega_2)
\end{eqnarray}

Consequently, the function $\cS \ni \omega \mapsto E(\omega)$
is convex. 

\subsection{Subadditivity of EoF}
To discuss this property, which seems 
to be important in quantum information (cf. \cite{Hor1}), 
we consider the tensor product of von Neumann algebras
$\cB(\cH_1) \otimes \cM \otimes \cB(\cH_1) \otimes \cM$
and a state $\omega \otimes \omega$ over it where $\omega$ is a state on
$\cB(\cH_1) \otimes \cM$.
We observe
\begin{eqnarray}
E(\omega \otimes  \omega) =
\inf_{\mu \in M_{\omega \otimes \omega}(\cS_T)}
\int d\mu(\nu) S_{1+2}(r\nu)  \le \nonumber \\ 
\inf_{\mu_1 \times \mu_2 \in M_{\omega}(\cS) \times M_{\omega}(\cS)}
\int d\mu_1(\nu) \int d\mu_2(\nu^{\prime}) S_{1+2}(r
\circ \nu \otimes  \nu^{\prime})
\nonumber \\
\le \inf_{\mu_1 \times \mu_2 \in M_{\omega}(\cS) \times M_{\omega}(\cS)}
\int d\mu_1(\nu) \int d\mu_2(\nu^{\prime}) (S_1(r\nu) \nonumber \\
+ S_1(r\nu^{\prime})) = 2 E(\omega)
\end{eqnarray}

where $\cS_T$ denotes the set of all states on 
$\cB(\cH_1) \otimes \cM \otimes \cB(\cH_1) \otimes \cM$,
$S_{1+2}$ ($S_1$) the von Neumann entropy on $\cB(\cH_1)
\otimes \cB(\cH_1)$ ($\cB(\cH_1)$ respectively).
The last inequality follows from subadditivity of the von Neumann 
entropy.
Consequently, EoF has also a form of subadditivity. 
Applying the above argument to $E(\omega \otimes ...\otimes \omega)$
one can consider the "density" of EoF and treat $E(\omega)$ as an extensive
(thermodynamical) quantity.

\subsection{Continuity of EoF}
As entanglement of formation, EoF, is a convex, real-valued function on the 
topological space $\cS$

\begin{equation}
\cS \ni \omega \mapsto E(\omega) \in \Rn
\end{equation}
it is natural to pose a question about its continuity.
Going in that direction we proved  (see \cite{M})
\begin{prop}
EoF, $\cS \ni \omega \mapsto E(\omega)$, is a continuous function.
\end{prop}
This result has the following important corollary. Namely,
as $\cS \ni \omega \mapsto E(\omega)$ is a continuous convex
function,  
an application of the Bauer maximum principle 
leads to:
\begin{cor}
$E(\omega)$ attains its maximum at an extremal point
of $\cS$, so the family of maximally entangled states is a subset
of pure states.
\end{cor}

\subsection{Comparison with the Bennett, DiVincenzo, Smolin and Wooters 
definition of EoF}
As our definition of EoF is a generalization of that given by Bennett {\it et al} 
(cf \cite{Ben}), it is natural to compare these
two definitions. Let us denote Bennett's {\it et al} entanglement of formation
by $EoF_B$. It is an easy observation that
$EoF \le EoF_B$. To examine
the converse inequality
we start with another simple observation that
\begin{eqnarray}
\inf_{\mu \in M_{\omega}(\cS)} \int d\mu(\nu) S(r\nu) \qquad \qquad \qquad \\
= \inf\{ \sum_{i=1}^n \lambda_i S(r\nu_i):    
\omega = \sum_{i=1}^n \lambda_i \nu_i \quad(convex\quad sum) \}
\nonumber\\
\end{eqnarray}
where the first infimum is attained for some $\mu \in M_{\omega}(\cS)$. 
The above 
observation follows from the fact that each measure $\mu$ can be ($^*$weakly) 
approximated by measures with finite support. 
On the other hand, measures concentrated on $\cS_p$,
where $\cS_p$ is the set of all pure states, are known to be
maximal with respect to the order $\mu \prec \nu$ ($\mu \prec \nu$
if and only if $\mu(f) \le \nu(f)$ for any convex, 
real-valued convex function $f$,
cf. \cite{phel} or \cite{Alf}), so minimal on the set of all concave
functions. It particular, such the measure
is minimal on $S\circ r$. Thus to get the converse
inequality, $EoF \ge EoF_B$ it would be enough to prove existence
of very special type of decompositions, so called
optimal decompositions.
A decomposition $\omega = \sum_{j=1}^n \lambda_j \varrho_j$, where
$\{ \varrho_i \}$ are pure states, such that the 
infimum in the definition of EoF is attained will be called 
{\it an optimal decomposition}. In other words, the infimum
is attained by a measure $\mu_0$ with finite support
contained in the set of all pure states. Thus, we want to have
$$E(\omega) = \inf_{\mu \in M_{\omega}(\cal S)}\int_{\cal S}
S(r \varrho) d\mu(\varrho) = \int_{\cal S} d\mu_0(\varrho) S(r \varrho)$$
with $supp \mu_0 = \{ \varrho_1,..., \varrho_n \}$, $n< \infty$
and $\varrho_i \in \cS_p$.
Here, $\mu_0 = \sum_1^n \lambda_i \delta_{\varrho_i}$
where $\delta_{\varrho}$ stands for the Dirac measure, $\{ \varrho_i \}$ are 
pure states and $\omega = \sum \lambda_i\varrho_i$.
In (\cite{M}) we proved:
\begin{prop}
The maximum of the set $\{ \mu(-S \circ r); \mu \in M_{\omega}(\cS) \}$
 for a continuous convex function $-S$ is attained
by a simplicial boundary measure.
\end{prop}
Then a straightforward application of the classical Carath\'eodory theorem 
(cf \cite{Alf}) leads to
\begin{cor} Assume that both Hilbert spaces $\cH_1$ and $\cH_2$ are finite dimensional.
Then, there exist optimal decompositions.
Therefore, our definition of EoF and that given by Bennett {\it et al} are equal to each other.
However, this is not true if the assumption on dimensionality of Hilbert spaces be dropped.
\end{cor}
\vskip 0.5cm
\section{Quantum correlations}
In this Section we introduce the notion of coefficient of quantum correlations
and we will look more closely at relations between 
quantum correlations and entanglement. We wish to start with a generalization
of the framework of the previous Sections. 
Let $\cA = \otimes_1^N \cA_i$ be a (e.g. spatial) tensor product of $C^*$-algebras $\cA_i$.
We assume that each $\cA_i$ contains the identity $\idty$.
Let $\phi$ be a state on $\cA$. 
Again, the set of all states on $\cA$ will be denoted by $\cS(\cA)$.
The pair $(\cA, \phi)$ will be 
considered as a (quantum) probability system.
Further, let $(a_1,...a_m)$ be a system of elements of $\cA$ such that for 
every $\nu = 1,2,..., m$ there is $i_{\nu} \in \{1,...,m\}$
such that $a_{\nu} \in \cA_{i_{\nu}}.$ To measure any correlations 
of the system we have to analyze the evaluation of a state $\phi$
on $m$-points $a_1,...,a_m$, i.e., $\phi(a_1,...,a_m)$.

In the sequel, considering $\phi(a_1,...,a_m)$, we will always assume
that $a_i \in \cA_i$ and indices are ordered. This is legitimate since
each $\cA_i$ can be embeded in $\cA$ and then the tensor product structure
implies that $a_i$ commutes with $a_j$ for $i \ne j$, $a_i \in \cA_i$, $a_j \in \cA_j$.
Consequently, we will consider $\phi(a_{{\nu}_1},..., a_{{\nu}_l})$ where
$(\nu_1,...,\nu_l) \subset \{1,...,N \}$ is an ordered subset and
$a_{{\nu}_i} \in \cA_{{\nu}_i}$.

Let us define, now in more general context, 
{\it the restriction map r} (cf. \cite{BR}). Let $\cB_1$ and $\cB_2$ be $C^*$-subalgebras
of the $C^*$-algebra $\cA$. Assume that $\cB_1$ and $\cB_2$ and $\cA$ have a common
identity, $\cB_1 \subseteq \cB_2^{\prime}$ and $\cB_1 \cup \cB_2$ generates $\cA$ as a 
$C^*$-algebra.
 Define the map $r:  \cS(\cA) \mapsto \cS(\cB_1)$ by
\begin{equation}
\label{rest}
(r\om)(a) = \om(a) \quad for \quad all \quad a \in \cB_1
\end{equation}
Specializing this definition for the tensor structure of $\cA = \otimes_{i=1}^N \cA_i$
one has the following definition 
 $r_{{\nu}_k} : \cS(\cA) \to \cS(\cA_{\nu_k})$, 
\begin{equation}
(r_{{\nu}_k}\om)(a) = \om(  \idty \otimes ... \otimes \idty \otimes \underbrace{a}_{{\nu}_k}
 \otimes \idty \otimes
... \otimes \idty) 
\end{equation}
Clearly, $r_{\nu_k}$ is an affine, $w^*$-continuous, ``onto", map. 
Again, as $\cS(\cA)$ is $^*$-weak 
convex compact set one can employ the Choquet theory. 
To this end, we denote by $M_{\omega}(\cS(\cA))$
the set of all positive, normalized Radon measures on $\cS(\cA)$ with the barycenter $\omega$.
Futhermore, we denote by
$M^0_{r_{{\nu}_k}\phi}(\cS(\cA_{{\nu}_k})) \subset 
M_{r_{{\nu}_k}\phi}(\cS(\cA_{{\nu}_k}))$
the set of all finitely supported positive normalized Radon measures.
Thus, if $\mu_{\nu_k}$ is in 
$M^0_{r_{{\nu}_k}\phi}(\cS(\cA_{{\nu}_k}))$, then
$\mu_{\nu_k} = \sum_1^P \la_i \delta_{\varrho_i^{\nu_k}}$
with $\sum_1^P \la_i \varrho_i^{\nu_k} = r_{\nu_k}\om$.
Again, $\de_{\varrho}$, stands for the Dirac (or point) measure.

Turning to quantum correlations, we recall that the entanglement 
is often considered as a signature of quantum correlations.
Although, the concept of quantum correlations is essential one for quantum statistical
mechanics, there is still lack of its precise definition.
To make an attempt to formulate a rigorous definition of quantum
correlations, guided by the (classical) probability theory with its 
definition of coefficient of independence, we will define the coefficient 
of quantum correlations. Leaving aside for a moment the general framework, let us 
present the basic idea for the simplest composite system, i.e.
a system consisting of two subsystems only. Thus, $\cA = \cA_1 \otimes \cA_2$. 
We note
\begin{rem}
\label{key}
Let us consider a separable state $\omega$ on $\cA \equiv \cA_1 \otimes \cA_2$,
$\omega(\cdot) \equiv Tr\{ (\sum_i \la_i \varrho^1_i
\otimes \varrho_i^2) \cdot \}$ and observe that, in general, 
$\omega(a \otimes \idty \cdot \idty \otimes b) \ne
\omega(a \otimes \idty) \omega(\idty \otimes b)$ for $a \in \cA_1$ and $b \in \cA_2$.
Thus, the state $\omega$ reflects some correlations. However, 
as the state $\omega$ is separable, these correlations are considered to be 
of classical nature only. Namely, each (classical) probability measure can be ($^*$-weakly)
approximated by a net of probability measures with finite support. Hence, each (classical)
probability measure on a composite system exhibits the basic properties of a separable state. 
\end{rem}

Therefore, to define ``pure'' quantum correlations we should ``substract'' classical correlations.
Suppose that a measure $\mu$ is in $M_{\phi}^0(\cS(\cA))$. So, $\mu = 
\sum_{i=1}^P \la_i \de_{\varrho_i}$ and the corresponding decomposition
of $\phi$ is given by $\phi = \sum_{i=1}^P \la_i \varrho_i$.
As $r_{1}$ ($r_2$) is an affine map of  $\cS(\cA)$ onto $\cS(\cA_1)$ ($\cS(\cA_2)$ respectively)
one has
\begin{equation}
r_{1} \phi = \sum_{i=1}^P \la_i \cdot r_{1} (\varrho_i)
\end{equation}
and
\begin{equation}
r_{2} \phi = \sum_{i=1}^P \la_i \cdot r_{2} (\varrho_i)
\end{equation}

Consequently, the decomposition of $\phi$ determined by $\mu$
induces the corresponding decomposition of $r_{1} \phi$ and $r_{2} \phi$
(determined by $\mu_1 = \sum_i \la_i \de_{r_{1} \varrho_i}$
and $\mu_2 = \sum_i \la_i \de_{r_{2} \varrho_i}$ respectively). 
More generally, let us define $\mu_I$ $(\mu_{II})$ on Borel subsets $F_I \subset \cS(\cA_1)$ 
($ F_{II} \subset \cS(\cA_2)$ respectively) by

\begin{equation}
\label{pobraz1}
\mu_{I} ( F_{I} ) = \mu (r^{-1}_{1} (F_{I})
\end{equation}
and
\begin{equation}
\label{pobraz2}
\mu_{II} ( F_{II} ) = \mu (r^{-1}_{2} (F_{II}) 
\end{equation}
where $\mu$ is a measure in $M_{\phi}(\cS(\cA)).$
This can be done as for any Borel set $F$
(for example, take as $F$ the subset $F_I \subset \cS(\cA_1)$), 
$r^{-1}(F)$ is a Borel set in $\cS(\cA)$.
Suppose that $F^0_I$ is a Borel subset in $\cS(\cA_1)$ such that
$F^0_I \supset \{r_1\rho_1, ..., r_1\rho_P \}$ and consider 
$\mu_I(\{F^0_I\}) \equiv \mu(r^{-1}_1(\{F^0_I \}))$, $\mu \in M_{\phi}(\cS)$. 
Clearly, $ F^0 \equiv r_1^{-1}(\{F^0_I \}) 
\supset \{ \rho_1,..., \rho_P \}$. But, if $\mu$ is supported by the subset
$\{\rho_1,...,\rho_P \}$ of $F^0$ then $\mu_I$ is supported on
$\{r_1\rho_1,\cdot \cdot \cdot, r_1\rho_P \}$.
Furthermore, assuming $\mu \in M^0_{\phi}(\cS(\cA))$ and noting
$r_1\phi = \int r_1 \xi d\mu(\xi) = \int \xi_r d\mu \circ r^{-1} (\xi_r)$
one has $\int\xi d\mu_I(\xi) = r_1 \phi$. Here, we denoted $r_1\xi$ by $\xi_r$.
Clearly, the same argument can be applied for $r_2$ and $\mu_{II}$.
In particular, one can easily note that
\begin{equation}
\label{gwiazdka}
\mu_I(\{ r_1 \rho_i \}) = \mu_{II}( \{ r_2 \rho_i \} ), \qquad i=1,...,P
\end{equation}
for any $\mu \in M^0_{\phi}(\cS)$.
Having measures $\mu_I$ on $\cS(\cA_1)$ and $\mu_{II}$ on $\cS(\cA_2)$, both originating 
from the measure $\mu$ on $\cS(\cA)$, we wish to define a new measure
$\boxtimes \mu$ on $\cS(\cA_1) \times \cS(\cA_2)$ which encodes classical 
correlations between two subsystems described by $\cA_1$ and $\cA_2$ respectively. 
As the first step we define it for discrete measures.
Let $\mu^d \in M^0_{\phi}(\cS)$, i.e. $\mu^d = \sum_i \la^d_i \delta_{\rho^d_i}$
with $\la^d_i \ge 0$, $\sum_i \la^d_i =1$, $\rho^d_i \in \cS(\cA)$.
Then, the just given argument leads to
$\mu^d_I = \sum_i \la^d_i \delta_{r_1\rho^d_i}$
and $\mu^d_{II} = \sum_i \la^d_i \delta_{r_2\rho^d_i}$. Define
\begin{equation}
\label{gwiazdka2}
\boxtimes \mu = \sum_i \la^d_i \delta_{r_1 \rho^d_i} \times 
\delta_{r_2 \rho^d_i}
\end{equation}
where we have used (\ref{gwiazdka}).
Now, let us take an arbitrary measure $\mu$ in $M_{\phi}(\cS)$. Then, there exists
net $\mu_k$ such that $\mu_k \in M^0_{\phi}(\cS)$ and $\mu_k \to \mu$ ($^*$-weakly).
Defining $\mu^k_I$ ($\mu_{II}^k$) analogously as $\mu_I$ ($\mu_{II}$
respectively; cf (\ref{pobraz1})) one has $\mu^k_I \to \mu_I$ and $\mu^k_{II} \to \mu_{II}$ where the convergence
is taken in $^*$-weak topology. Then define, for each $k$,
$\boxtimes \mu^k$ as in (\ref{gwiazdka2}). 
One can verify that $\{ \boxtimes \mu^k \}_k$
is convergent to a measure on $\cS(\cA_1) \times \cS(\cA_2)$,
so taking the weak limit we arrive
to the measure $\boxtimes \mu$ on $\cS(\cA_1) \times \cS(\cA_2)$.
All that leads to

\begin{defn}
\begin{enumerate}
\item
Let $\cA$ be a $C^*$-algebra with two $W^*$-subalgebras  
$\cB_1$, $\cB_2$ satisfying conditions given
prior to formula (\ref{rest}) suplemented by the condition $\cB_2 \subset \cB_1^{\prime}$.
The coefficient of quantum correlations for the state $\phi$ evaluated on $a_1a_2$, 
$\phi(a_1a_2)$, $a_i \in \cB_i$, $i=1,2$, is defined as
\begin{equation}
CQC(\phi;a_1,a_2) = \inf_{\mu \in M_{\phi}(\cS(\cA))}
|\Bigl( \int \xi d ( \mu)(\xi) \Bigr) (a_1a_2)
 - \Bigl( \int \xi d (\boxtimes \mu)(\xi) \Bigr) (a_1a_2)|
\end{equation}
\begin{equation}
\equiv \inf_{\mu \in M_{\phi}(\cS(\cA))}
|\phi(a_1a_2) - \Bigl( \int \xi d (\boxtimes \mu)(\xi) \Bigr) (a_1a_2)|
\end{equation}
where $a_i \in \cB_i$, $i=1,2.$
\item
Assume that $\cA = \cA_1 \otimes \cA_2$. Then, specializing
the definition of CQC to the tensor structure of the $C^*$-algebra $\cA$ we have
the degree of quantum correlations for the state $\phi$ evaluated on $a_1\otimes a_2$.
It is defined as
\begin{equation}
d(\phi; a_1, a_2) = \inf_{ \mu \in M_{\phi}(\cS(\cA))} |\phi(a_1 \otimes a_2) -
\Bigl(\int \xi d(\boxtimes \mu)(\xi)\Bigr)(a_1 \otimes a_2)|
\end{equation}
where $a_i \in \cA_i$, $i=1,2.$
\end{enumerate}
\end{defn}

To comment on the above definition, firstly we note that the definition of $CQC$ makes
no appeal to the tensor structure of $\cA$. Therefore, such a definition seems to be 
very well adapted to the general theory of quasi-local algebras, so to the general theory 
of quantum systems (cf. \cite{Haag}). Clearly, we followed the classical
definition of coefficient of independence. The main difference between the classical and 
quantum approaches (apart from the existence of extra correlations) relies on the fact
that  in the quantum setting, the set of states does not form a simplex. Therefore, 
there is no uniquness in decomposition of a (quantum) state. Hence, to carry out 
our construction we are forced to take the infimum operation over the set of 
all ``good'' decompositions.

Secondly, to have the framework well adapted to an analysis of separable (so also 
entangled) states it is necessary to take into account the tensor structure of the
algebra. To distinguish these two cases, we give two different names to a measure of 
quantum correlations: coefficient (degree respectively) of quantum correlation.

Turning to separable states we have

\begin{prop}
A state $\phi$ is separable one if and only if $d(\phi;a_1,a_2) = 0$
for any $a_1,a_2$.
\end{prop}
$\pr$
Recall that
$\phi$ is separable iff $EoF(\phi)=0$ (cf. Section III).
Hence, there exists a measure $\mu^0$ such that $\phi = \int \xi d\mu^0 (\xi)$
with the property that $supp\mu$ is contained in the following set
$\overline{conv} \{ \xi_1 \otimes \xi_2; \xi_1 \in \cS(\cA_1), \xi_2 \in \cS(\cA_2) \}$.
An application of the restriction maps $r_{1}$ and $r_2$ to the measure $\mu^0$
lead to measures $\mu^0_I$
and $\mu^0_{II}$.
Then, considering the ($^*$-weak) approximation one has $\mu^0 = lim \mu^0_k$ with
$\mu^0_k = \sum \la^k_i \delta_{\phi^I_{i,k}} \times \delta_{\phi^{II}_{i,k}}$
where $\phi^a_{i,k} \in \cS(\cA_a)$, $a \in \{ 1,2 \}$. Clearly, $\mu^0_{\alpha,k} \equiv
\mu^0_k \circ r_a =\sum \la^k_i \delta_{\phi^{\alpha}_{i,k}}$, $\alpha \in \{I,II \}$.
Therefore, $\mu^0_k = \boxtimes \mu^0_k$.  Hence,
$d(\phi;a_1,a_2)$ is equal to $0$.
Conversely, suppose that $d(\phi;a_1,a_2) = 0$ for any $a_1, a_2$.
As $M_{\phi}(\cS(\cA))$ is compact, then $inf$ in definition of 
$d(\phi;a_1,a_2)$ is attainable.
Therefore, there exist
two measures $\mu_I$ and $\mu_{II}$ defining $\boxtimes \mu$
such that
\begin{equation}
\phi(a_1 \otimes a_2) = \bigl(\int \xi d(\boxtimes \mu)(\xi)\bigr)(a_1 \otimes a_2)
\end{equation}
However, this proves the separability.
\QED

The Proposition may be summarized by saying that any separable state contains 
classical correlations only. Therefore, an entangled state contains ``non-classical''
(or quantum) correlations.

\begin{rem}
$CQC$ yields information about quantum correlations and therefore it makes
legitimate to apply $CQC$ for an analysis of quantum stochastic dynamics.
However, this topic exceeds the scope of this paper and it will be present in 
another paper (see \cite{KM2}, and \cite{KM})
\end{rem}

Turning to the general case, $\cA = \otimes_{i = 1}^N \cA_i$, 
to each state $\phi$ on $\cA$
we will assign the family of product states
\begin{equation}
\label{1}
\{ \phi^{\la}_{{\nu}_1} \otimes ... \otimes \phi^{\la}_{{\nu}_{\la}}
 \}_{\la \in \Lambda}
\end{equation}

where for each $\la \in \Lambda$, $\phi^{\la}_{{\nu}_k} \in supp 
\mu_k$ for some $\mu_k \in
M_{r_{{\nu}_k}\phi}(\cS(\cA_{{\nu}_k}))$. We recall that
$M_{r_{{\nu}_k}\phi}(\cS(\cA_{{\nu}_k}))$ stands for all
all normalized positive Radon measures with barycenter of the restricted state
$r_{{\nu}_k} \phi$ on $\cA_{{\nu}_k}$.
Let us define
\begin{equation}
 \overline{conv} 
 \{ \phi^{\la}_{{\nu}_1} \otimes...\otimes \phi^{\la}_{{\nu}_l}
 \}_{\la \in \Lambda} \equiv \cS_{cc}
\end{equation}
where, by a slight abuse of notation we denote an extension
of $\phi^{\la}_{{\nu}_1} \otimes...\otimes \phi^{\la}_{{\nu}_l}$ to a state over $\cA$
by the same letter.

We have observed that one can interpret 
a state in $\cS_{cc}$ as a state encoding classical 
correlations only. Therefore, a state in $\cS_{cc}$ will be called a {\it c-dependent state}.
As $CQC$ measures the deviation of correlations of a state from classical correlations,
going in that direction, we propose

\begin{defn}
\label{dwa}
Let $\cA = \otimes_{i=1}^N \cA_i$ and
let a state $\phi$ be in $\cS(\cA)$. Then
\begin{enumerate}
\item
\begin{equation}
d(\phi, \cS_{cc}) = \inf_{\psi \in \cS_{cc}} || \phi - \psi||
\end{equation}
will be called the uniform degree of quantum correlation (UDQC).
\item
\begin{equation}
d_{\phi}(a_{\nu_1},...,a_{\nu_l}) = \inf_{\psi \in \cS_{cc}} |\phi(a_{\nu_1},...,a_{\nu_l})
- \psi(a_{\nu_1},...,a_{\nu_l})|
\end{equation}
will be called the weak degree of quantum correlation (WDQC).
 Here, we recall that $(\nu_1,...,\nu_l) \subset \{1,...,N \}$ is an ordered subset.
\end{enumerate}
\end{defn}
\smallskip
We close this Section with some remarks on Definition \ref{dwa}.
Firstly, it is an easy observation that $d(\phi,\cS_{cc}) = 0$ if and only
if $\phi$ is a separable state. Secondly, the equality
$d_{\phi}(a_1,...,a_m) =0$ can be treated as a definition of
{\it quantum independence} of subsystems of a composite system. 
However, we would like to emphasize that 
the subsystems still can have a ``classical'' correlations. 
Finally, let us specialize Definition 4.2 to a quantum chain, i.e.
$\cA = \otimes_{i \in \Zn} M_d(\Cn) \equiv \cA_{\Zn}$, $\cA_1 = \cA_{(- \infty, 0)}$,
$\cA_2 = \cA_{\{1 \}}$, ..., $\cA_{N-1} = \cA_{\{ N-2 \}}$, $\cA_N = \cA_{ (N-1, \infty)}$.
Here, the algebra $M_d(\Cn)$ associated with each site $i$ is taken to be the full 
algebra of $d \times d$ matrices.
Then, the subset of states with $WDQC > 0$ can be called finitely quantum correlated states
(cf \cite{Bruno}).

\end{document}